\documentclass[pra,twocolumn,floatfix]{revtex4}

\newcommand{\beq}{\begin{equation}}
\newcommand{\eeq}{\end{equation}}
\newcommand{\ket}[1]{\ensuremath{\left| {#1} \right>}}
\newcommand{\bra}[1]{\ensuremath{\left< {#1} \right|}}
\newcommand{\braket}[2]{\ensuremath{\left< \left. {#1} \right| {#2} \right>}}

\usepackage{graphicx}
\usepackage{amsmath}
\usepackage{amssymb}

\begin{document}
%\draft

\title{Laser linewidth effects in quantum state discrimination by EIT}

\author{M. J. McDonnell, D. N. Stacey and A. M. Steane
%\address{
\\
\small Centre for Quantum Computation, Department of Atomic and Laser Physics, \\
\small Clarendon Laboratory, Parks Road, Oxford, OX1 3PU, England}

\date{\today}

\begin{abstract}
We discuss the use of electromagnetically modified
absorption to achieve selective excitation in atoms: that is, the
laser excitation of one transition while avoiding simultaneously
exciting another transition whose frequency is the same as or
close to that of the first. The selectivity
which can be achieved in the presence of electromagnetically
induced transparency (EIT) is limited by the decoherence rate of
the dark state. We present exact
analytical expressions for this effect, and also physical models
and approximate expressions
which give useful insights into the phenomena. 
When the laser frequencies are near-resonant with the single-photon atomic
transitions, EIT is essential for achieving discrimination. 
When the laser frequencies are far detuned, 
the `bright' two-photon Raman resonance is important
for achieving selective excitation, while the `dark' resonance
(EIT) need not be. The application to laser cooling of a trapped atom is
also discussed.
\end{abstract}

\maketitle

%\twocolumn

Electromagnetically-induced transparency (EIT, also called dark resonance or
coherent population trapping) and phenomena related to it have
been widely studied (see for example \cite{arimondo1996,book:scully} and references therein).
These two-photon resonance phenomena can give rise to sharp spectral
features, which can be used for various purposes, including for example
magnetometry and laser cooling
\cite{fleischhauer1994,fleischhauer1995,aspect1988,lindberg1986,morigi2000,gea1995}.  
Recently, it was shown that EIT could be used to allow the angular
momentum state of an atom to be detected with high quantum efficiency
even in the absence of a Zeeman effect (i.e. at zero applied magnetic field and/or
zero magnetic dipole moment of the atom) \cite{mcdonnell2003a}. This paper develops the
theory relevant to the latter, and sheds light on related experimental techniques
such as laser cooling.

The essential concept here is the use of a two-photon resonance
to achieve selective excitation in atoms. 
We are concerned with two states, generally closely-spaced, which have allowed transitions
separated in frequency by a small interval (or coincident in frequency). 
We denote these
states by $\ket{S}$ and $\ket{I}$, for `suppressed' and `interacting' respectively.
Let $P_S$, $P_I$ be an experimentally observed signal, such as
collected fluorescence, obtained when the atom is prepared in
$\ket{S}$ or  $\ket{I}$, respectively. We wish to irradiate the atom in such a way as to
achieve a detectable signal $P_I$ and maximise the ratio $r \equiv P_I / P_S$. 

The states $\ket{S}$ and $\ket{I}$ could for example
be magnetic substates of the same atomic energy level, or
they could represent the same internal
state, but different motional states of an atom, such as
two vibrational states in a harmonic potential well.
In the former case, a high value for $r$ permits the atomic spin state
to be detected \cite{mcdonnell2003a}; in the latter, a high value for
$r$ implies that efficient laser cooling is possible
\cite{lindberg1986,morigi2000,marzoli1994,reiss2002}. 

Suppose the signal is collected fluorescence.
Excitation out of a state $\ket{S}$ can sometimes be avoided by using
light of appropriate 
polarization. For example, with circularly polarized light driving the transition
$n {\rm s}\, ^2{\rm S}_{1/2}$--$n {\rm p}\, ^2{\rm P}_{1/2}$, one of the 
$^2{\rm S}_{1/2}$ magnetic sublevels does not couple to the radiation. This would allow
$P_S \simeq 0$ (limited only by experimental precision).
However, in such a
case the population of $\ket{I}$ is rapidly lost by optical pumping to $\ket{S}$,
and hence $P_I$ is also small. Our interest
here is in achieving high values of $r$ without significant transfer of population between
$\ket{I}$ and $\ket{S}$.

The basic idea of using EIT, and more generally electromagnetically modified
absorption (EMA), is illustrated in figure 1. We consider two situations.
In the case illustrated in figure 1a, both $\ket{S}$
and $\ket{I}$ are connected by strong (e.g. electric-dipole
allowed) transitions to upper states, such that the two
transition frequencies are close together or even identical,
but $\ket{S}$ is part of a three-level manifold D which can
exhibit dark resonance, while $\ket{I}$ is not. In the case
illustrated in figure 1b, both $\ket{S}$ and $\ket{I}$ are
each part of separate three-level manifolds (called D and B for
`dark' and `bright' respectively); both manifolds are driven
simultaneously by a single pair of laser
beams. 

Suppose the detected signal were the fluorescence from the atom. 
In either case (a) or (b), if the
laser frequencies are chosen in such a way that the
D manifold is at a dark resonance, but the B
manifold is not, then in the limit of no decoherence of the dark state,
the ratio $r \rightarrow \infty$. This is evident when the manifolds D and
B are not connected, since then excitation from $\ket{S}$ will stop once
the atom spontaneously enters the dark state, while excitation from $\ket{I}$
can continue indefinitely. It is also true when the upper state of manifold
D can decay to $\ket{I}$ (which is more usual in practice), as long
as we ensure an atom
prepared in $\ket{S}$ remains dark as the laser beams are introduced.
This can be done by introducing the `pump' laser, Rabi frequency $\Omega_2$ in
figure 1, first, and then switching on the `probe' laser of Rabi frequency $\Omega_1$
adiabatically,
i.e. on a time-scale slow compared to the light-shift caused by the pump laser.

In practice the
available value of $r$ is therefore limited by the
loss of coherence of the dark state.  For brevity we
refer to this loss of coherence as a laser linewidth effect,
although it can also be caused by other mechanisms.
It is modelled simply as a decay
rate $\gamma$ of the off-diagonal density matrix
element $\rho_{21}$ in the Optical Bloch Equations for the D manifold.
Note that many studies of phenomena related to EIT do not need to
take this decoherence rate into account, except as a refinement, but
here it is central.

\begin{figure}[t]
% \psfrag{Bstate}{\ket{I}}
% \psfrag{Dstate}{\ket{S}}
% \psfrag{om1}{$\Omega_{1}$}
% \psfrag{om2}{$\Omega_{2}$}
% \psfrag{gam1}{$\Gamma_{1}$}
% \psfrag{gam2}{$\Gamma_{2}$}
% \psfrag{gamlw}{$\gamma$}
\centerline{\resizebox{8 cm}{!}{\includegraphics{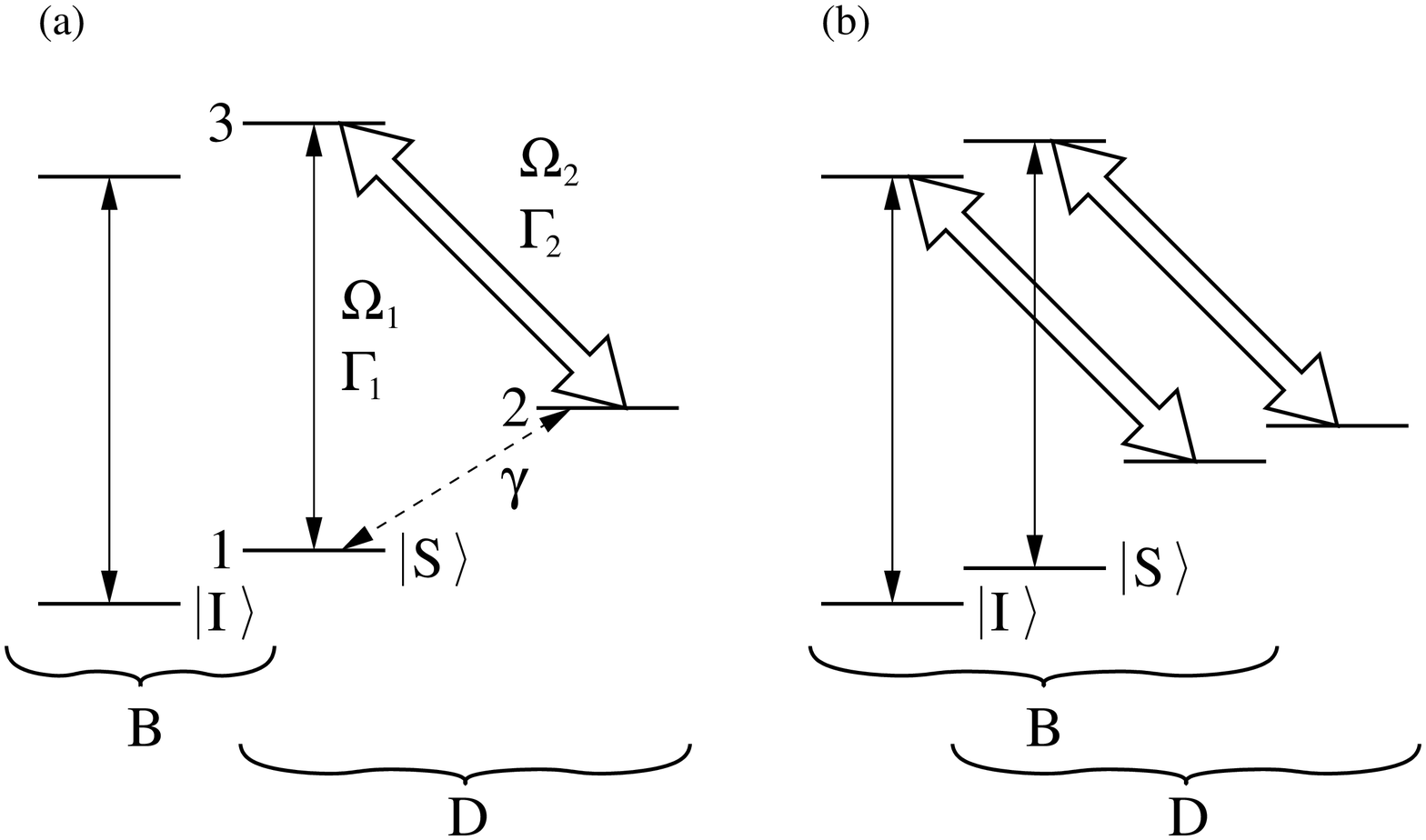}}}
\center
\rule{0pt}{24 pt}
\caption{Atomic level schemes considered in the text. 
a: $\ket{I}$ is
part of a two-level manifold, $\ket{S}$ is part of
a three-level manifold. b: $\ket{I}$ and $\ket{S}$ are
each part of separate three-level manifolds.
In either case, the atom is illuminated by a single
pair of laser beams which drive both manifolds; 
the single photon
transition 1--3 in the D manifold is either degenerate with or close to
the single-photon transition out of $\ket{I}$ in the B manifold. 
$\Omega_1$, $\Omega_2$ are Rabi frequencies, $\Gamma_1$,
$\Gamma_2$ spontaneous decay rates, $\gamma$ is the rate
of decay of coherence between levels 2 and 1.
Both types of energy level structure are common in groups of
atomic levels with $J \ne 0$. Case (b) also occurs in
the combination of
internal and vibrational states of a trapped atom.} \label{f:1}
\end{figure}

We wish to understand the selectivity $r$ which can be achieved, as a function
of all the relevant parameters. In order to do this, we model the atom as
if the two manifolds $B$ and $D$ were not connected. If the 
excited state of $D$ can in fact decay to $B$ then such a model remains
a good approximation as long as
the population of the excited state of $D$ is small. It will be seen that
this is the case when $r \gg 1$. On the other hand, if the excited state of
$B$ can decay to $D$ then the model does not apply. (In any case
this situation would result in
optical pumping from $\ket{I}$ to $\ket{S}$ and hence only a small signal
$P_I$.)

We assume the experimental signals $P_S$ and $P_I$ are proportional to the steady-state
population of the excited state in the relevant manifold. This ignores a possible
contribution from the initial transient
behaviour, for example during adiabatic switching on of the laser beams.
The ignored contribution is negligible when the time-scale on which the measured
signal is obtained is long compared with the transient. 

Our approach is to write down the steady state solution to the optical
Bloch equations (OBEs) for a three-level atom excited by
two laser fields of finite linewidth, and then examine the behaviour of this solution.
The full solution is a rather complicated function of many parameters. In previous
work it has been obtained and then studied in a simplified form under various
restrictions, such as low pump power or zero detuning. 
One of the aims of
this paper is to provide analytical expressions which retain 
as great a range of validity as possible, while being sufficiently
simple to give clear general insights into the physical behaviour. 
This is done by finding factorisations of parts of the formulae, 
and by making good choices of the parameters with which
to express them. We also present physical pictures to give further
insight into the behaviour. 

The work was motivated by the idea that the phenomenon of
dark resonance ought to make available especially high
values of $r$. Our results show, however, that this is only
partially true.

We consider two regimes in detail: first
the resonant case $\Delta_1 = \Delta_2 = 0$,
and then the far-detuned case $\Delta_1 \gg \Gamma$
where $\Delta_1 = \omega_{L1} - \omega_{31}$, 
$\Delta_2 = \omega_{L2} - \omega_{32}$ are the detunings of
the lasers from their respective single-photon transitions, and $\Gamma$ is
the width of the upper state.

The case of figure 1a is interesting because it permits
a high degree of state discrimination even when the
single-photon transitions from $\ket{S}$ and $\ket{I}$
have the same frequency. In this situation frequency
discrimination of the bare single-photon transitions is ruled
out completely, hence the EMA is crucial to achieving any
discrimination. It was shown in ref \cite{mcdonnell2003a}
that this can be used
to measure an atomic spin state
at zero magnetic field or zero magnetic dipole moment.
The choice $\Delta_1 = \Delta_2$ is used to make
the dark resonance of the $D$ system as dark
as possible, while setting both detunings equal to zero causes
the $B$ system to give the maximum
single-photon scattering rate. The value of $r$ is derived
in section \ref{s:zero}; it is found to be proportional
to the intensity of the pump laser in the D system, divided
by $\gamma$.

In the case of figure 1b, both manifolds D and B exhibit
the phenomena of dark and bright 2-photon resonances.
In order to obtain a good discrimination at finite laser
linewidth, we require a frequency separation between
the bright resonances of the two manifolds.
This will occur either if there are suitable
energy level separations in the atomic structure,
or if the coupling strengths
on the pump transitions are sufficiently different to
cause a substantial difference in a.c. Stark shifts (light
shifts) in the two manifolds.
We discuss the case of figure 1b in detail because
it is more complicated and the results are surprising.
We find that although tuning the D manifold to dark
resonance does not do any harm (for the purpose
of maximising $r$), it does not permit
any increase in the value of $r$ compared to that available at large
$\Delta_1, \Delta_2$, 
where the dark resonance disappears.
Furthermore, the fact that the dark resonance causes
one side of the Fano profile to fall substantially
below a Lorentzian profile of the same height and width,
which suggests that it would enhance
discrimination, is misleading.
It turns out that at given laser linewidth, the best choice
of the other laser parameters is such that
the width of the Fano profile is dominated by
the laser linewidth, and in this situation it takes a Lorentzian form.

These conclusions apply when the decoherence of the
dark state is caused by phase diffusion, leading to
Lorenztian lineshapes. When other noise
sources dominate, such as laser drift
or jitter with a non-Lorentzian profile, then
the presence of a dark resonance can, in contrast, be
useful.  

In the context of laser cooling, the implication is that
for given laser intensities and linewidths, 
the intrinsic lower limit on  the  steady-state temperature is always obtained at
large detuning, where the bright resonance is important but
the dark resonance (EIT) is not.  
However, when further heating mechanisms are present the
dark resonance may be useful since it provides an increased
cooling rate for a given temperature.

The paper is organized as follows. Section \ref{s:single}
briefly presents the case of frequency discrimination
using single-photon excitation, in order to have
a performance measure with which to compare our results.
Section \ref{s:OBE} presents the OBEs
and their steady-state solution. Section \ref{s:zero} discusses
the resonant case $\Delta_1 = \Delta_2 = 0$,
and section \ref{s:far} discusses the far-detuned case
$\Delta_1 \gg \Gamma$. We simplify
the equations and
present two physical models which give useful insights
into the bright resonance and its dependence on
the laser parameters. Section \ref{s:bright} then
discusses the discrimination which is available
by using the bright resonance in the situation
of fig. 1b. In section \ref{s:cool} the same ideas are applied
to the case of laser cooling of a trapped atom or ion,
by presenting numerical solutions of the master equation describing
the evolution of both internal and motional states, in the
Lamb-Dicke limit.

\section{Narrow single-photon transitions} \label{s:single}

Before examining the 2-photon phenomena, we use
a simpler situation to provide a `benchmark' with which
to compare the performance. Suppose the states $\ket{S}$ and
$\ket{I}$ were each part of a closed 2-level manifold, both with a
long-lived upper state, so that the excitation linewidth is dominated by
laser linewidth. We can then obtain selective excitation by
using a single laser beam tuned to resonance with the $B$
manifold. The excitation rate as a function of laser frequency is
Lorentzian, with FWHM given by the laser linewidth $\gamma_{\rm L}$. When
system $B$ is resonant, system $D$ is driven off-resonantly, with
detuning $Z$ given by the separation of the two transitions
involved. We assume the atom-laser coupling (e.g. the electric dipole
matrix elements) to be the same for the two transitions. Then
the Lorentzian excitation profile gives the excitation ratio
\beq
  r = \left( \frac{ 2 Z }{\gamma_{\rm L}} \right)^2  + 1 . \label{r1}
\eeq

\section{Optical Bloch Equations for 3-level atom} \label{s:OBE}

We adopt an interaction picture. Then in 
the rotating wave approximation (RWA),  
the OBEs for a 3-level $\Lambda$ system with
two lasers are: (c.f. \cite{stalgies1998,whitley1976})

\begin{widetext}
\begin{eqnarray}
\dot{\rho}_{33} &=& -\Gamma \rho_{33} - i(\rho_{13}-\rho_{31}) \Omega_1 / 2 - i (\rho_{23} - \rho_{32}) \Omega_2 / 2, \label{1}\\
\dot{\rho}_{11} &=&  \Gamma_1 \rho_{33} + i(\rho_{13}-\rho_{31}) \Omega_1 /2, \\
\dot{\rho}_{22} &=&  \Gamma_2 \rho_{33} + i(\rho_{23}-\rho_{32}) \Omega_2 /2, \label{3} \\
\dot{\rho}_{13} &=&  (-\Gamma_{13} - i \Delta_1) \rho_{13} - i(\rho_{33}-\rho_{11}) \Omega_1 /2 + i \rho_{12} \Omega_2/2, \\
\dot{\rho}_{23} &=&  (-\Gamma_{23} - i \Delta_2) \rho_{23} - i(\rho_{33}-\rho_{22}) \Omega_2 /2 + i \rho_{21} \Omega_1/2, \\
\dot{\rho}_{12} &=&  i(\Delta_2 - \Delta_1) \rho_{12} + i\rho_{13}\Omega_2 /2 - i \rho_{32} \Omega_1/2 - \gamma
\rho_{12},   \label{6}
\end{eqnarray}
\end{widetext}

where $\Omega_{1}$ and $\Omega_{2}$ are the Rabi frequencies of the
`probe' and `pump' lasers exciting transitions 1--3 and 2--3 respectively,
$\Gamma$ is the decay rate of the
upper state 3, $\Gamma_1$ and $\Gamma_2$  
are the decay rates of 3 to 1 and 2 respectively (in a closed system, $\Gamma
 = \Gamma_1 + \Gamma_2$); 
the decay rates of the coherences are
$\Gamma_{13},\Gamma_{23}, \gamma \equiv \Gamma_{12}$. These can all be
independent quantities. However, in the case where the coherence
decay is purely associated with the finite lifetime of level 3,
and with laser linewidths $\gamma_1, \gamma_2$, the coherence
decay rates are given by
\begin{eqnarray}
\Gamma_{13} &=& (\Gamma + \gamma_1)/2 \\
\Gamma_{23} &=& (\Gamma + \gamma_2)/2  \label{gamma23} \\
\gamma &=& (\gamma_1 + \gamma_2)/2.  \label{addgamma}
\end{eqnarray}
The last equation, (\ref{addgamma}), applies when the two laser
beams have independent dephasing, which is typically the case
if they originate in different lasers. If they both originate
in the same laser, with a frequency difference imposed by
another device such as an acousto-optic modulator, then
(\ref{addgamma}) does not apply and instead
$\gamma$ is equal to the rate of dephasing of the
imposed frequency difference.

Any one of (\ref{1}) to (\ref{3}) can be replaced using the
normalisation condition
  \beq \rho_{11} + \rho_{22} + \rho_{33} = 1   \label{norm}
\eeq
in order to get a linearly independent set of equations.
The general solution of (\ref{1})--(\ref{norm}) in steady state
is given in the appendix.

We define a parameter $\alpha \equiv 2 \Gamma_{13} / \Gamma$.
The definition implies that $\alpha \simeq 1$ when $\gamma \ll \Gamma$.
In the rest of the paper, we will make the simplifying assumption
$\Gamma_{23} = \Gamma_{13}$, so that both are equal to
$\alpha \Gamma /2$. This is valid when the lasers
linewidths are equal, and approximately valid when they
are unequal but small compared to $\Gamma$.  When $\Gamma_{13}
= \Gamma_{23}$, the steady state value
of the upper state population is
\beq
\rho_{33} = 2 \Omega_1^2 \Omega_2^2
\frac{\left[ 2 \alpha \Gamma ( \delta^2 + \gamma^2 )
+ (\Omega_1^2 + \Omega_2^2) \gamma \right]}
{c_0 + c_1 \gamma + c_2 \gamma^2}         \label{rho33exact}
\eeq
where $\delta = \Delta_1 - \Delta_2$ is the detuning from
the dark resonance condition, and the coefficients
$c_i$ in the denominator are given in the appendix. 

\begin{figure}[t]
\centerline{\resizebox{!}{6 cm}{\includegraphics{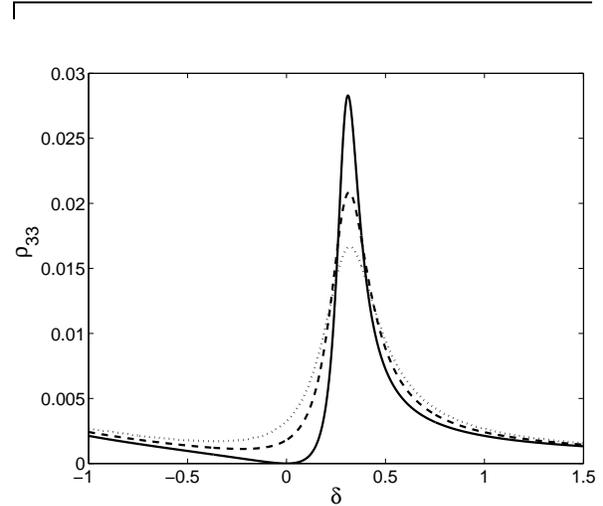}}}
\caption{Example fluorescence profiles for a set of values
of laser linewidth. The parameter values are (in units where
$\Gamma=1$):
$\Omega_1=0.1$, $\Omega_2=1$, $\Delta_2 = 3$, $\Gamma_1 = \Gamma_2
= 0.5$, and $\gamma=0,0.05,0.1$ for full, dashed, dotted curves,
respectively. Note that at finite $\gamma$ 
the absorption minimum is displaced with respect
to $\delta=0$, as remarked by Kofman \cite{kofman1997}.
} \label{f:2}
\end{figure}

Some example profiles of the 2-photon resonance, as described by equation
(\ref{rho33exact}), are shown in figure \ref{f:2}. This illustrates
the change in shape of the resonance as $\gamma$ increases.

Although it is useful to have the full expression (\ref{rho33exact}),
it is too unwieldy to yield simple insights into the behaviour.
We therefore examine it in two limiting cases.

\section{Resonant lasers}  \label{s:zero}

In the situation shown in figure 1a, and such that the lower and upper
energy levels in the B manifold are degenerate with states 1 and 3 (respectively)
in the D manifold, then
in order to optimize the discrimination
factor $r$ we choose  $\Delta_{1} = \Delta_{2}  =  0$. There is then
a dark resonance in the D manifold, while the B manifold is at a maximum in the
fluorescence rate. The absorption in the D manifold is not
completely cancelled owing to a non-zero decoherence rate $\gamma$.

For both lasers on resonance with their respective transitions, 
a factor $(\Omega_1^2 + \Omega_2^2 + 4 \Gamma_{13} \gamma)$ cancels
in the full expression (\ref{rho33exact}) for the excited
state population in the D manifold. The expression reduces to 
\beq
\rho_{33}^{\rm D} = \cfrac{2\gamma\Omega_{1}^{2}\Omega_{2}^{2}}
  {\Omega^{2}Y+2\gamma\left(3\Omega_{1}^{2}\Omega_{2}^{2}+2\Gamma_{13}Y\right)}         
\label{rho33darkres}
\eeq
where $\Omega^{2}\equiv\Omega_{1}^{2}+\Omega_{2}^{2}$ and
$Y\equiv\Gamma_{2}\Omega_{1}^{2}+\Gamma_{1}\Omega_{2}^{2}$. 

Assuming the atom--laser coupling constants are such that the Rabi frequency
in the B manifold is equal to $C \Omega_1$, where $C$ is a constant
(such as a Clebsch-Gordan coefficient, for example), and that the
excited state in B has the same total decay rate $\Gamma$ as the
excited state in D, then
the excited state population for the (two level) B manifold is 
\beq
\rho^{\rm B}=\cfrac{1}{2+\Gamma^{2}/C^2 \Omega_{1}^{2}} \, .
\eeq

The ratio of steady-state populations is therefore
\beq
r=\cfrac{\rho^{\rm B}}{\rho_{33}^{\rm D}}=
\cfrac{\Omega^{2}Y+2\gamma\left(3\Omega_{1}^{2}\Omega_{2}^{2}+2\Gamma_{13}Y\right)}
{2\gamma\Omega_{2}^{2}\left(2\Omega_{1}^{2}+\Gamma^{2}/C^2 \right)} \, .  \label{ratio0}
\eeq
This result is valid without restriction---no assumptions have yet been made about
the laser intensities or atomic parameters (except those implicit in a master equation
treatment in RWA).

In the limit of low probe laser intensity compared to the pump laser intensity, i.e.
\beq
\Omega_{1}^{2}\ll\Omega_{2}^{2},\;\cfrac{\Gamma_{1}}{\Gamma_{2}}\Omega_{2}^{2}\, ,
\eeq
the ratio is
% \begin{eqnarray}
%   r & = & \cfrac{ \left(\Omega_{2}^{2}+4\gamma \Gamma_{13}\right) \Gamma_1 C^2}
%     {2\gamma\Gamma^2} \\
%   & \approx & \cfrac{\Omega_{2}^{2} \Gamma_1 C^2}{2\gamma\Gamma^2} \, .
% \end{eqnarray}
\begin{eqnarray}
  r & \approx & \cfrac{\Omega_{2}^{2}\Gamma_{1}}
    {2\gamma\left(2\Omega_{1}^{2}+\Gamma^2/C^{2}\right)} \\
  & \approx & \left\{
    \begin{array}{lr}
      \cfrac{\Omega_{2}^{2} \Gamma_1 C^2}{2\gamma\Gamma^2} & 
      ,\;\;\;\Omega_{1}^{2}\ll\Gamma^{2}, \\
      \cfrac{\Omega_{2}^{2} \Gamma_1}{4\gamma\Omega_{1}^2} & 
      ,\;\;\;\Omega_{1}^{2}\gg\Gamma^{2} \, .
    \end{array}
    \right.
\end{eqnarray}
% Hence for example when $C \sim 1$ and $\Gamma_1 \sim \Gamma$, then
% very good discrimination efficiency is obtained if
% $\Omega_{2}^{2}\gg\gamma\Gamma$. 
Hence a large enough pump laser intensity permits very good discrimination
to be achieved.

Figure 3 shows the steady state populations in the excited state for
the $\ket{I}$ and $\ket{S}$ systems with the pump laser at zero detuning $\Delta_2=0$,
as a function of the probe laser detuning $\Delta_1$. The example parameter
values are chosen to illustrate a case where $\ket{I}$ and $\ket{S}$ are adjacent
Zeeman sublevels in the atomic ground state 
at zero magnetic field, and the excited states decay primarily
to the ground state; c.f. low-lying levels in
alkaline earth ions, as discussed in \cite{mcdonnell2003a}.

\begin{figure}[t]
\centerline{\resizebox{!}{6 cm}{\includegraphics{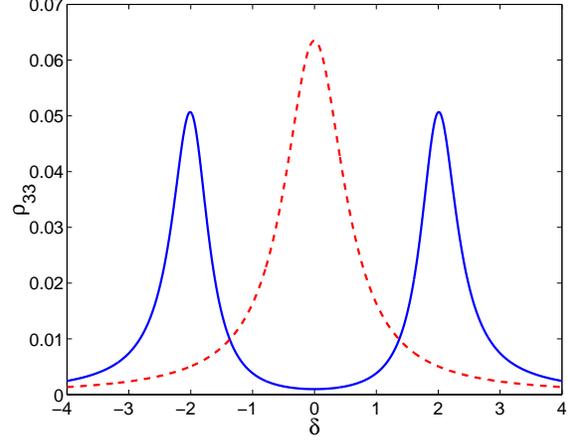}}}
\caption{Example of state discrimation for atomic structure
of the form shown in figure 1a. The curves show the steady-state
value of the excited state population
for an atom prepared in $\ket{S}$ (full
curve) and $\ket{I}$ (dashed curve), respectively, as a function
of detuning $\delta$. 
The B manifold shows the standard `2-level atom' Lorentzian profile,
while the D manifold shows a dark resonance at
$\Delta_1=0$ in between two peaks at $\pm\Omega_{2}/2$  (these 
show the positions of the dressed states created by
the pump laser). By choosing $\delta = 0$ the ratio of excitation rates
is maximised.  The parameter values for the three level system are (in
units where $\Gamma=1$): 
$\Omega_{1}=0.2$, $\Omega_{2}=4$, $\Gamma_{1}=\Gamma_{2}=0.5$,
$\gamma=0.1$.  For the two level
system $\Omega_{1}=0.2\sqrt{2}$,
$\Omega_{2}=0$.}
 \label{f:3}
\end{figure}

In the case of a ladder system, i.e. when level 2 lies above level 3 in
the D manifold, the results are as follows. The OBEs, eq. (\ref{1}) and (\ref{3})
are modified so that the spontaneous emission at rate $\Gamma_2$ is now from
2 to 3, not the other way around. The steady state solution at zero detuning is
\begin{widetext}
\beq
\rho_{33} = \frac{2 \gamma \Omega_1^2 \Omega_2^2 + \Omega_1^2 \Gamma_2 \left(
\Omega_1^2 + 4 \Gamma_{23} \gamma \right) }
{ \Omega^2 \tilde{Y} - \Gamma_2 \Omega_1^2 \left( 3 \Omega_2^2 + 2\Gamma_1
 \left(\Gamma_{23}-\Gamma_{13}\right)\right)
  + 2 \gamma 
  \left(3 \Omega_1^2 \Omega_2^2 + 2 \Gamma_{13} \tilde{Y} + 4 \Gamma_2
  \left(\Gamma_{23}-\Gamma_{13}\right)\Omega_1^2\right) },
 \label{rholadder}
\eeq
\end{widetext}
%\rho_{33} = \frac{2 \gamma \Omega_1^2 \Omega_2^2 + \Omega_1^2 \Gamma_2 \left(
%\Omega_1^2 + 4 \Gamma_{23} \gamma \right) }
%{ \Omega^2 \tilde{Y} - \Gamma_2 \Omega_1^2 \left( 3 \Omega_2^2 + \Gamma_1 \Gamma_2 \right)
%+ 2 \gamma 
%\left(3 \Omega_1^2 \Omega_2^2 + 2 \Gamma_{13} \tilde{Y} + 2 \Gamma_2^2 \Omega_1^2\right) },
%\label{rholadder}
where $\tilde{Y} = 2 \Gamma_2 \Omega_1^2 + \Gamma_1 \Omega_2^2 + 2 \Gamma_1 \Gamma_2 \Gamma_{23}$.
%and we assumed $\Gamma_{23} - \Gamma_{12} = \Gamma_2/2$ in order to simplify the expression.
In the case where the coherence decay rates are purely due to spontaneous emission and laser
linewidths, then for the ladder system, equations
(\ref{gamma23}) and (\ref{addgamma}) should be replaced by
\begin{eqnarray}
\Gamma_{23} &=& (\Gamma + \Gamma_2 + \gamma_2)/2 \\
\gamma &=& (\Gamma_2 + \gamma_1 + \gamma_2) / 2.
\end{eqnarray}
(In a closed system, $\Gamma=\Gamma_1$).
%Hence the assumption $\Gamma_{32} - \Gamma_{12} = \Gamma_2/2$ is valid when the laser
%linewidths are equal, and is approximately valid when they are unequal but
%small compared to $\Gamma$.
In the limit $\Omega_2^2 \gg \Omega_1^2$ expressions (\ref{rholadder}) and (\ref{rho33darkres})
are the same.

\section{Weak probe, large detuning} \label{s:far}

We next examine the behaviour for a weak probe intensity and large detunings:
\begin{eqnarray}
\Omega_1^2 &\ll& \frac{\Gamma_1}{\Gamma_2}\Omega_2^2,\; \Gamma_1 \alpha \Gamma\, ,   \label{app1} \\
\Delta_1^2 &\gg& \alpha^2 \Gamma^2,\; \delta^2  \label{app2}\, .
\end{eqnarray}

Under the weak probe condition (\ref{app1}) alone (i.e. without any restriction
on detunings), we obtain
\begin{eqnarray}
c_0 &\simeq& 16 \Omega_2^2 \Gamma_1 \left[ \rule{0 pt}{2.5 ex}
\Delta_1^2 ( \delta - \Delta' )^2
+ \delta^2 (\alpha \Gamma/2)^2 \right.\nonumber\\
&& \left. + \delta^2 \Delta_2^2  \frac{\Omega_1^2}{\Omega_2^2} \frac{\Gamma_2}{\Gamma_1}
 + \frac{\Omega_2^2 \Omega_1^2}{16}
\left( \frac{\Gamma_2}{\Gamma_1} + 2 \right) \right]     \label{c0}\\
c_1 &\simeq& 16 \Omega_2^2 \Gamma_1
\left[ \frac{\alpha \Gamma\Omega_2^2}{4} + \frac{ \Omega_1^2 }{2 \alpha \Gamma \Gamma_1}
(\Gamma_1 \Delta_1^2 + \Gamma_2 \Delta_2^2 \right. \nonumber \\
& & \left.+ (\Gamma_1 + \Gamma_2) \Delta_1 \Delta_2) \rule{0 pt}{2.5 ex} \right]
       \label{c1}    \\
c_2 &\simeq& 16 \Omega_2^2 \Gamma_1 (\Delta_1^2 + \alpha^2 \Gamma^2/4)  \label{c2}
\end{eqnarray}
where
\beq
\Delta'\equiv\frac{\Omega_2^2}{4 \Delta_1}.
\eeq
When $\Delta_1 = \Delta_2$, then $\Delta'$ is the light shift of
the states 2 (upwards when $\Delta_2 > 0$)
and 3 (downwards when $\Delta_2 > 0$)
caused by the pump laser.

If condition (\ref{app2}) applies,
there is a further simplification of the expressions for $c_i$, and
substituting them into (\ref{rho33exact}) gives
\begin{widetext}
\beq
\rho_{33} = \frac{\Omega_{\rm eff}^2 \left(
\alpha \left(\frac{\delta^2 +
\gamma^2}{2 \Delta'^2}\right) R + \gamma \right) / 2 \Gamma_1}
{(\delta-\Delta')^2 +
\left( \frac{\delta \Gamma}{2 \Delta_1} \right)^2
\left( \alpha^2 +
\frac{\Gamma_2}{\Gamma_1} \frac{\Omega_{\rm eff}^2}{R^2} \right)
+ \frac{\Omega_{\rm eff}^2}{4} \left( \frac{\Gamma_2}{\Gamma_1} + 2
\right)
+ \left( \alpha + \frac{\Omega_{\rm
eff}^2}{R^2}\frac{\Gamma}{\alpha  \Gamma_1} \right) R
\gamma  + \gamma^2 }                        \label{full}
\eeq
\end{widetext}
where
\beq
R \equiv \frac{\Omega_2^2}{4 \Delta_1^2} \Gamma         \label{R}
\eeq
is (when $\Delta_1 \simeq \Delta_2$) the scattering rate on the
strongly driven transition 2--3 per unit population in 2, and
\beq
\Omega_{\rm eff} \equiv \frac{\Omega_2 \Omega_1}{2 \Delta_1}     \label{Reff}
\eeq
is the effective Rabi frequency for Rabi flopping on the Raman resonance
between levels 1 and 2. The reason for introducing $R$ and
$\Omega_{\rm eff}$ is that they yield physical insights which
will become apparent below.

Many previous treatments of this problem in the limit
(\ref{app1}) have assumed the further condition $\Omega_2
\gg \Omega_1 \Delta_1 / \Gamma$, which may usefully be written
$R \gg \Omega_{\rm eff}$. It will be important for some
of the results to be discussed that we have not made this
assumption. A nice feature is that we can find
readily understandable physical pictures for this more
general case.

The fact that we have not assumed $R \gg \Omega_{\rm eff}$
implies that our results remain valid
at large $\Delta_1$. For example,
away from the 2-photon resonance, i.e. $|\delta| \gg
|\Delta'|$, the terms proportional to $\delta^2$ in (\ref{full})
dominate, and the result is
\beq
\rho_{33} \rightarrow \frac{\Omega_1^2 \Gamma/\Gamma_1}{4
\Delta_1^2}.         \label{single}
\eeq
This agrees with the prediction of the rate equations for
the three-level system. It can be understood as the excited state
population due to single-photon excitation from level 1 by the weaker
laser, with the stronger
laser playing the role of `repumper'.

At small $\Delta_1,\;\Delta_2$ equation (\ref{full}) remains fairly
accurate for small laser linewidth, since the terms which were neglected
under assumption (\ref{app2}) are primarily in $c_1$ and $c_2$, not $c_0$.

\subsection{Zero laser linewidth}

Let us consider the situation at zero laser linewidth, in order
to obtain some physical insights.
In this case, $\gamma = 0$
and $\alpha = 1$. Equation (\ref{full}) simplifies
to
\begin{widetext}
\beq
\rho_{33} = \frac{\Omega_1^2 \delta^2  \Gamma / \Gamma_1}
{4 \Delta_1^2 ( \delta - \Delta' )^2
+ \delta^2 \Gamma^2
+ 4 \delta^2 \Delta_1^2
\frac{\Omega_1^2}{\Omega_2^2} \frac{\Gamma_2}{\Gamma_1}
+ \frac{\Omega_2^2 \Omega_1^2}{4}
\left( \frac{\Gamma_2}{\Gamma_1} + 2 \right)}   \label{simp}
\eeq
\end{widetext}
(We present the equation in terms of $\Omega_2$, $\Omega_1$
and $\Delta_1$ in order to facilitate comparison with previous
work \cite{stalgies1998,leibfried2003}.) This has a zero at $\delta=0$
(the dark resonance) and a peak 
at $\delta \simeq \Delta'$ (the bright resonance). The precise location
of the peak is discussed in \cite{stalgies1998}.

The denominator of (\ref{simp}) can be simplified to good
approximation by replacing the occurrences
of $\delta^2$ by $\Delta'^2$ while retaining the $(\delta-\Delta')$
term. This is a good approximation because it is accurate
when $\delta = \Delta'$, and away from this detuning,
the first term in the denominator dominates when $\Delta_1$ is
large. This substitution gives the canonical `Fano' type of profile
\cite{Fano1961}:
\beq
\rho_{33} \simeq \frac{\Omega_{\rm eff}^2  \left(  \delta / \Delta'
\right)^2  R/4 \Gamma_1}
{ (\delta-\Delta')^2  + R^2 / 4
+  \Omega_{\rm eff}^2 \Gamma/2 \Gamma_1}.      \label{rhoapp}
\eeq
The width of the peak is now
easy to extract. The values of $\delta$ at which
$\rho_{33}$ is half its maximum value are given by
\beq
(\delta - \Delta') \simeq \frac{f}{2} \left( \frac{f}{\Delta'} \pm
1 \right).      \label{dhalf}
\eeq
where
\beq
 f = \left( R^2 +
\Omega_{\rm eff}^2 \frac{2 \Gamma}{\Gamma_1} \right)^{1/2}    \label{FWHM}
\eeq
is the FWHM of the peak
and to simplify the RHS we used the condition $f \ll \Delta'$
(which follows from (\ref{app1}),(\ref{app2})).

We next present some physical insights into the behaviour.

\subsubsection{Two models}

The main features of $\rho_{33}$ are the zero at dark resonance
and the peak at the bright resonance. 

As many authors have discussed \cite{arimondo1996,book:scully}, the zero is
due to a cancellation between the two
excitation routes when the atomic state is $(\Omega_2 \ket{1} -
\Omega_1 \ket{2}) (\Omega_1^2 + \Omega_2^2)^{-1/2}$ in
an interaction picture. When
$\delta = 0$ this is a stationary state,
so once in it the atom does not precess out of it.

To understand the bright resonance, we present two physical models.
The first is the well-known `dressed atom' approach; the second is an
alternative model based on Rabi flopping and the quantum Zeno effect.
For general reviews and references on the quantum Zeno effect, see
for example refs \cite{misra1977,beige1996,power1996,misra2003}.

\begin{figure}[t]
% \psfrag{omeff}{\LARGE{$\Omega_{\mbox{eff}}$}}
% \psfrag{3n1}{\LARGE{\ket{\tilde{3},n-1}}}
% \psfrag{2n}{\LARGE{\ket{\tilde{2},n}}}
% \psfrag{3n}{\LARGE{\ket{\tilde{3},n}}}
% \psfrag{2n1}{\LARGE{\ket{\tilde{2},n+1}}}
% \psfrag{om1}{\LARGE{$\Omega_{1}$}}
% \psfrag{R}{\LARGE{$R$}}
\centerline{\resizebox{8 cm}{!}{\includegraphics{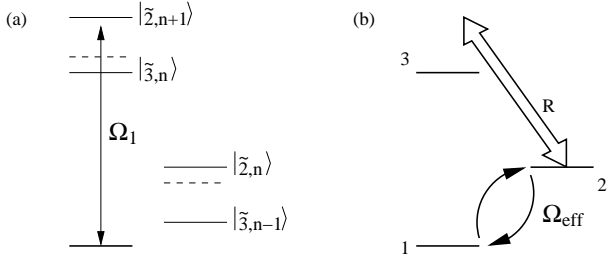}}}
\caption{Physical models of the bright resonance. a: The pump
laser dresses the atom, the probe laser excites the atom from
$\ket{1}$ to the dressed states. These give a broad resonance
displaced by $-\Delta'$ and a narrow resonance displaced
by $\Delta_2 + \Delta'$ from the position of the undressed
excited state $\ket{3}$. b: The two lasers together drive
Rabi flopping between (undressed) levels $1$ and $2$
by a Raman transition,
and the pump laser off-resonantly excites transitions from $\ket{2}$ to
$\ket{3}$. The Rabi flopping is resonant when the laser
frequencies match the energy difference between 1 and
the Stark-shifted level 2.} \label{f:4}
\end{figure}

The application of the `dressed atom' treatment to EIT and related phenomena
has been widely discussed; see
\cite{arimondo1996,book:scully} for an introduction and further references.
In this model, the behaviour may be regarded as one in which the probe
laser excites population from level 1 to a dressed state created by
the intense pump laser (see figure \ref{f:4}a).
Near the centre of the bright resonance, i.e.
when $\delta \simeq \Delta'$, eq. (\ref{rhoapp}) takes the form
\beq
\rho_{33} \simeq \frac{\Omega_{\rm eff}^2 R/4 \Gamma_1}
{ (\delta-\Delta')^2  + R^2 / 4
+  \Omega_{\rm eff}^2 \Gamma/2 \Gamma_1}.   \label{rhodress}
\eeq
Comparing this with the well-known expression for the upper state population
of a two-level atom in steady state, it is seen that the result has
a natural interpretation in the dressed atom model. The 
dressed state has decay rate $R$ and the strength of the coupling to it
is $\Omega_{\rm eff}$. The two terms which make up the FWHM (\ref{FWHM})
of the resonance are then to be interpreted as `natural linewidth' and
`power broadening' of the dressed state.

Our alternative model is
based on Rabi flopping and the Zeno effect, as follows (c.f. \cite{jong1997}).

When the difference frequency $\delta$ is tuned to the light shift
$\Delta'$, the pump and probe lasers drive resonant Rabi flopping
between level $1$ and the light-shifted level $2$.
Observe that when $\Omega_2 \gg \Omega_1$ and neither of the single-photon
transitions
are saturated, the population $\rho_{33}$ is produced primarily by
excitation from level $2$. The excited state population thus
comes about from the combination of the Rabi flopping which moves
population between $1$ and $2$, and single-photon
excitation from $2$ to $3$ (see figure \ref{f:4}b).
However, the single-photon
excitation results in a spontaneously emitted photon
when $3$ decays, and therefore constitutes a
measurement of the atom's state
in the $1$, $2$ basis. This measurement suppresses
the Rabi flopping by the quantum Zeno effect.
The steady state solution
finds a balance between these effects.

This physical picture suggests the following analysis. We take
the limit  $\Omega_2 \gg \Omega_1$ such that population in 3
is produced purely by excitation from 2 by the pump laser,
and treat this by the rate equation
\beq
\dot{\rho}_{33} = R_2 \rho_{22} - \Gamma \rho_{33}
\eeq
where the single-photon excitation rate $R_2$ is given by the
Fermi Golden Rule:
$
R_2 = (\pi/2) \Omega_2^2 g(\Delta_2)
$
where $g$ is a lineshape function.
Hence in steady state,
\beq
\rho_{33} = \frac{R_2 \rho_{22}}{\Gamma}.     \label{rho33}
\eeq
The spontaneous decay of $\rho_{33}$
leads to a Lorentzian lineshape of width $\Gamma$, so in the limit
$\Delta_2 \gg \Gamma,\, \Omega_2$,
\beq
R_2 \simeq \frac{\Omega_2^2}{4 \Delta_2^2} \Gamma.
\eeq

We calculate the steady-state population $\rho_{22}$ by considering the Rabi
flopping between levels 1 and 2, and taking $\rho_{22}$ to be the
mean population averaged over time. When $R_2$ is
sufficiently small, and the Raman process is resonant,
this Rabi flopping leads to equal average populations $\rho_{11}$ and
$\rho_{22}$, i.e. both equal to $1/2$. When $R_2$ is
non-negligible, on the other hand, the Rabi flopping is
interrupted by photon scattering events. These act like
measurements, and suppress the flopping by the Zeno effect
when they are sufficiently frequent.

An uninterrupted Rabi flopping process would cause the population $\rho_{22}$ to vary
with time as:
\beq
\rho_{22}(t) = \frac{\Omega_{\rm eff}^2}{\delta'^2 + \Omega_{\rm eff}^2} \sin^2 \frac{1}{2}
(\Omega_{\rm eff}^2 + \delta'^2)^{1/2} t                    \label{flop}
\eeq
where $\delta' = \delta - \Delta'$ is the detuning from the Raman
resonance (bright resonance), $\Omega_{\rm eff}$ is given in
equation (\ref{Reff}), and we assumed the initial condition
$\rho_{22} = 0$ for convenience (but we expect that the mean
population to be calculated will not depend on the initial
conditions). The photon scattering acts both as a measurement-type
process, collapsing the state to either $1$ or $2$, and also causes
optical pumping to $1$. We will treat a simplified case in which
we assume the population always goes to $1$ after photon
scattering, and then the population in $2$ recommences evolving
as (\ref{flop}). This would be the behaviour to be expected
when $\Gamma_1 \gg \Gamma_2$. In this case the mean population of 2 is
estimated as
\beq
\bar{\rho}_{22} \simeq
\int_0^{\infty} P(t) \rho_{22}(t) dt   \label{mean}
\eeq
where $P(t) = R_2 e^{-R_2 t}$
is the probability that there is an interval $t$ between scattering
events. Performing the integral in (\ref{mean}) we obtain
\beq
\bar{\rho}_{22} \simeq \frac{1}{2} \frac{ \Omega_{\rm eff}^2 }{ \delta'^2  + R_2^2 + \Omega_{\rm eff}^2}
\eeq
and substituting this in (\ref{rho33}) gives
\beq
\rho_{33} \simeq \frac{ \Omega_{\rm eff}^2 R_2/2 \Gamma}
{ \delta'^2 + R_2^2 + \Omega_{\rm eff}^2}       \label{rhozeno}
\eeq
Note the similarity between equations (\ref{rhozeno}) and
(\ref{rhodress}). The Zeno effect calculation reproduces the
OBE result when $\Gamma_1 \simeq \Gamma$, except for
factors of 2 associated with $R_2$ and $\Omega_{\rm eff}^2$.
This confirms that it
gives a good physical insight into the behaviour. Of course a
full quantum Monte-Carlo type of calculation \cite{dalibard1992,plenio1998}
would reproduce the 
OBE result exactly. The present result simply
demonstrates the validity of the `Rabi-flopping/Zeno effect' physical
picture.

\subsubsection{Two regimes}

The above insights allow us to identify two distinct regimes of behaviour.
When $R \gg \Omega_{\rm eff}$, the Zeno
effect strongly suppresses the Rabi flopping. In this `Zeno
regime',
\beq
\rho_{33}^{\rm max} =\frac{\Omega_{\rm eff}^2}{R \Gamma_1} = \frac{\Omega_1^2}{\Gamma_1 \Gamma}, \;\;\;\;
{\rm FWHM} = R     \label{zenopeak}
\eeq
The interpretation in the dressed state picture is that of weak excitation, such
that the FWHM is equal to the dressed state's `natural linewidth'
$R$. 

When $R^2 \ll \Omega_{\rm eff}^2$ we obtain
\beq
\rho_{33}^{\rm max} =\frac{R}{2\Gamma}, \;\;\;\;
{\rm FWHM} =(2 \Gamma/\Gamma_1)^{1/2} \Omega_{\rm eff}.
\eeq
Here the Rabi flopping leads to $\rho_{11} =\rho_{33} \simeq 1/2$,
which leads directly to the value of $\rho_{33}^{\rm max}$, in
particular the fact that it depends purely on $R$.
The width of the resonance results from the
detuning-dependence of the Rabi flopping, and thus is governed
purely by $\Omega_{\rm eff}$. In the dressed state picture this
is the case where `power broadening' dominates.

\subsection{Finite laser linewidth}

We return to
equation (\ref{full}) in order to consider the effect
of finite laser linewidth. A useful approximation is the same
`trick' as was used for eq. (\ref{rhoapp}) where we replace
the $\delta^2$ in the denominator by $\Delta'^2$. This
considerably simplifies the denominator without much loss
of accuracy:
\begin{widetext}
\beq
\rho_{33} \simeq \frac{\Omega_{\rm eff}^2 \left(
\alpha  \left(\frac{\delta^2 + \gamma^2}{2\Delta'^2}\right) R
+ \gamma \right) / 2 \Gamma_1}
{(\delta-\Delta')^2 + ( \alpha R / 2 + \gamma )^2
+ \Omega_{\rm eff}^2 (\Gamma/2\Gamma_1) ( 1 + 2 \gamma / \alpha R )}
   \label{fullapp}
\eeq
\end{widetext}
Note that this result is similar to (\ref{rhoapp}) with
the substitution $R \rightarrow \alpha R + 2\gamma
\simeq R + 2 \gamma$. In other words, the
main effect of finite laser linewidth is to increase
the `linewidth' term in (\ref{rhoapp}) by $2 \gamma$.
In the Zeno picture this is an
illustration of the fact that measurement-induced collapses
have the same effect on a system as phase fluctuations.
Their effects add to produce the overall linewidth.

\subsubsection{Effect of laser linewidth on dark resonance}

The conditions (\ref{app1}), (\ref{app2}) imply $\Omega_{\rm eff}^2 \ll
\Delta'^2$. At $\delta=0$, this can be used to simplify the denominator
of (\ref{fullapp}). If we further assume
\beq
\gamma \ll \Omega_2^2 / \Gamma   \label{smallgam}
\eeq
(which is not a severe constraint on the range of validity of the
results) then we obtain
\begin{eqnarray}
\lefteqn{  \rho_{33}^{\rm dark} \simeq
\frac{\Omega_{\rm eff}^2  \gamma / 2 \Gamma_1 }
{\Delta'^2 + \Omega_1^2 \gamma/ \alpha \Gamma_1
+ \gamma^2}   }      \label{dark}  \\
&=& \frac{ 2 \Omega_1^2  \gamma / \Gamma_1 }
{\Omega_2^2 + (\Omega_{\rm eff}^2/R^2) (4 \Gamma^2/\alpha\Gamma_1) \gamma
+ (4 \Gamma/R) \gamma^2} .\nonumber \\ && \label{dark2}
\end{eqnarray}

\begin{figure}[t]
% \psfrag{gamt}{\LARGE{$\tilde{\Gamma}$}}
% \psfrag{gam1}{\LARGE{$\Gamma_{1}$}}
% \psfrag{gam2}{\LARGE{$\Gamma_{2}$}}
% \psfrag{+}{\LARGE{\ket{+}}}
% \psfrag{-}{\LARGE{\ket{-}}}
% \psfrag{3}{\LARGE{\ket{3}}}
% \psfrag{R}{\LARGE{$R$}}
\centerline{\resizebox{!}{4.5 cm}{\includegraphics{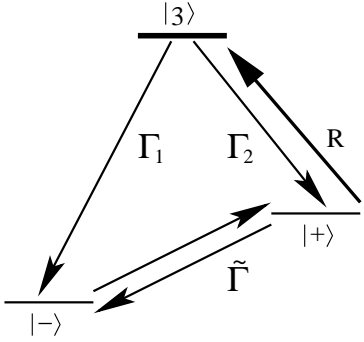}}}
\caption{Physical model of the effect of decoherence on
a dark resonance. The atom is analysed in the basis $\ket{3}$, $\ket{+}$,
$\ket{-}$, where $\ket{-}$ is the dark state. A simple rate
equation picture, with rates as shown, suffices to give the main
features of the behaviour.} \label{f:5}
\end{figure}

This result can be interpreted as follows.
The dark state is
\beq
\ket{-} = (\Omega_2 \ket{1} - \Omega_1 \ket{2}) (\Omega_1^2 +
\Omega_2^2)^{-1/2}.
\eeq
The combination of $\ket{1}$ and $\ket{2}$ that is orthogonal
to this is
\beq
\ket{+} = (\Omega_1 \ket{1} + \Omega_2 \ket{2}) (\Omega_1^2 +
\Omega_2^2)^{-1/2}.
\eeq
Decoherence associated with finite laser linewidth evolves
the state towards a random mixture of $\ket{-}$ with the
state $\ket{\sim}$  given by
\beq
\ket{\sim} = (\Omega_2 \ket{1} + \Omega_1 \ket{2}) (\Omega_1^2 +
\Omega_2^2)^{-1/2}.
\eeq
A good insight is obtained by analysing the system in the 
orthonormal basis
$\{ \ket{3},\;\ket{-},\;\ket{+} \}$ (see figure \ref{f:5}).
A complete master equation can be obtained in this basis \cite{arimondo1996}; that
of course gives exactly the same predictions as those given
by the OBEs in their standard form. However, it is noteworthy
that the dependence of $\rho_{33}$ on $\gamma$ at the
dark resonance point can be obtained
to second order in $\gamma$ by a rate equation approach, as follows.

The atom--laser interaction
Hamiltonian is $H_I = \Omega_1
\ket{3} \bra{1} + \Omega_2 \ket{3} \bra{2}$, and the only non-zero
matrix element of $H_I$ in the chosen basis is
$\bra{3} H_I \ket{+} = (\Omega_1^2 + \Omega_2^2)^{1/2}$.
When $\Omega_2^2 \gg \Omega_1^2$,
the spontaneous decay of $\ket{3}$ to $\ket{-}$ ($\ket{+}$) is at rate
approximately $\Gamma_1$ ($\Gamma_2$) respectively, owing to
the relative proportions of $\ket{1}$ and $\ket{2}$ in each of
$\ket{-}$ and $\ket{+}$. We model phase decoherence by
a spontaneous decay at the rate $\tilde{\Gamma}$
in both directions between $\ket{-}$ and $\ket{+}$.
The rate is given by the decay rate $\gamma/2$
between $\ket{-}$ and $\ket{\sim}$, multiplied by the
probability that an atom in $\ket{\sim}$ would be found in $\ket{+}$ if measured
in the $\ket{\pm}$ basis:
\beq
\tilde{\Gamma} = \frac{\gamma}{2} \left| \braket{\sim}{+} \right|^2
= \frac{2 \gamma \Omega_1^2 \Omega_2^2}{(\Omega_1^2 +
\Omega_2^2)^2}
\; \simeq \; 2 \gamma \Omega_1^2 / \Omega_2^2 \, .
\eeq
Invoking the limit (\ref{app2})
to simplify the atom-light coupling term, the resulting set of rate equations is
\begin{eqnarray}
\dot{\rho}_{33} &=& \left( \rho_{++} - \rho_{33} \right) R - \Gamma \rho_{33} \\
\dot{\rho}_{--} &=& \rho_{33} \Gamma_1 + \left( \rho_{++} - \rho_{--} \right)
\tilde{\Gamma} \\
1 &=& \rho_{33} +  \rho_{--} + \rho_{++}.
\end{eqnarray}

The solution is
\begin{eqnarray}
\rho_{33} &=& \frac{ R \tilde{\Gamma} } { R \Gamma_1 + 2 (R +
\Gamma) \tilde{\Gamma} } \\
&  \simeq &
\frac{2 \Omega_1^2  \gamma / \Gamma_1}
{\Omega_2^2 + (\Omega_{\rm eff}^2/R^2)(4 \Gamma^2/\Gamma_1)
\gamma }         \label{mydark}
\end{eqnarray}
where we used $\tilde{\Gamma} \ll \Gamma_1$ which follows
from (\ref{app1}).
Equation (\ref{mydark}) correctly reproduces all the features of (\ref{dark2})
up to second order in $\gamma$. The essence of the dynamics
when $\Gamma_1 \gg R \gg \tilde{\Gamma}$ is that population
moves from 3 to the dark state at the rate $\Gamma_1$, and from
the dark state to 3 (via $\ket{+}$) at the rate $\tilde{\Gamma}$.

Next we consider the overall shape of the 2-photon
resonance. The range of values of $\delta$ which
interests us is from $0$ to approximately $\Delta'$, the position
of the bright resonance. Examining (\ref{fullapp}) we
find that when
the laser linewidth is sufficient to
produce the condition
  \beq \gamma \gg \alpha  R   \label{Gcond} \eeq
then the
$\gamma$ term in the numerator dominates the other
terms.
In this case there is no longer
a local minimum near $\delta=0$; the dark resonance is completely
`washed out'. Therefore the condition (\ref{Gcond}) is sufficient
to change the overall lineshape to one close to a Lorentzian
function. Note that (\ref{Gcond})
always occurs at sufficiently large $\Delta_1$, independent of
the values of the other parameters.

In the case (\ref{Gcond}) and
when also $\gamma \gg \Omega_{\rm eff}, \Omega_{\rm eff}^2/R$,
the complete expression (\ref{full}) becomes simply
a Lorentzian function of linewidth $\gamma$,
for $|\delta| \le |\Delta'|$ (and for large $|\delta|$, see
eq. (\ref{single})).

\subsubsection{Effect of laser linewidth on bright resonance}

At the position of the bright resonance ($\delta = \Delta'$),
the condition (\ref{smallgam}) is sufficient to make the
$\gamma^2$ term in the numerator of (\ref{fullapp})
negligible. In this case (\ref{fullapp}) gives
\beq
\rho_{33}^{\rm bright} = \frac{ (\Omega_{\rm eff}^2 /2 \Gamma_1)
( \alpha R / 2 + \gamma )}
{ ( \alpha  R / 2 + \gamma )^2
+ \frac{1}{2} \Omega_{\rm eff}^2 (\Gamma/\Gamma_1) ( 1 + 2\gamma / \alpha R)
}  .  \label{bright}
\eeq
In the `Zeno regime' $\Omega_{\rm eff}^2 \ll R^2$ this leads to
the simple result
\beq
\rho_{33}^{\rm bright} \rightarrow
\frac{\Omega_{\rm eff}^2 / 2 \Gamma_1}{\alpha R/2 +
\gamma}.      \label{bright2}
\eeq

\section{Using the bright resonance for selective excitation}
\label{s:bright}

We will now explore the use of the bright resonance as a
sharp spectral feature, able to resolve two closely spaced
transitions. We have in mind the situation where the atomic
structure consists, to good approximation, of two $\Lambda$-
systems `side by side' as in fig. 1b. (Similar results can be expected
for two ladder-systems).
Each of the levels 1,2,3 are
split into two closely spaced components (such as Zeeman
sublevels, or two rungs of a ladder of vibrational energy levels). We still
have just two lasers, and we would like to drive one
$\Lambda$- system without driving the other.

The system we want to drive is B and the
system we would like not to drive is D. 
The measure of good discrimination to be adopted is the
ratio $r$ between the steady state value for $\rho_{33}$ in the
systems or manifolds B and D. 

In this section we will discuss the
case where the two manifolds
have the same coupling constants, so the same Rabi
frequencies $\Omega_2$, $\Omega_1$, but different energy level
spacings, such that when the Raman detuning is $\delta$ in system
B, it is $\delta - Z$ in system D. The discrimination ratio is then
\beq
r = \frac{\rho_{33}(\delta)}{\rho_{33}(\delta-Z)}  \label{rdef}
\eeq
where $\rho_{33}(\delta)$ is given by (\ref{full}).
The effect of a difference in coupling constants between the two
manifolds is discussed in the appendix.

First let us consider the case
$\Delta_1 \gg \Omega_1, \Omega_2, \gamma, \Gamma$, which we will
refer to for brevity as ``$\Delta_1 \rightarrow \infty$''.
This tells us the behaviour of $r$ at large
detunings. Equation (\ref{full}) gives:
\beq
\rho_{33}(\Delta_1 \rightarrow \infty)
= \frac{ \Omega_{\rm eff}^2 \left( 2 \alpha \Gamma (\delta^2 +
\gamma^2) / \Omega_2^2  + \gamma \right) / 2 \Gamma_1 }
{ (\delta - \Delta')^2 + \gamma \Omega_1^2 / \alpha \Gamma_1
+ \gamma^2 } .        \label{inf}
\eeq

At large $\Delta_1$, the light shift is small
compared to $Z$, so to produce the discrimination factor $r$
the dark resonance is irrelevant. We
tune system B to bright resonance, and it is found that
$r$ is maximised when
$\Omega_1^2 \ll \gamma \Gamma \ll \Omega_2^2$.
In this case, using (\ref{inf}) and (\ref{rdef}),
\beq
r (\Delta_1 \rightarrow \infty) = \frac{Z^2 + \gamma^2}
{\left(2 \alpha \Gamma Z^2 / \Omega_2^2 + \gamma \right) \gamma}.
 \label{rinf}
\eeq

Next let us consider the case where we arrange that $\Delta' = Z$.
This means that when the B system is tuned to bright resonance,
the D system is simultaneously tuned to dark resonance, and we
expect a large value for $r$. Examining the ratio
$r = \rho_{33}^{\rm bright} / \rho_{33}^{\rm dark}$ given by equations
(\ref{bright}) and (\ref{dark}), it
is found that $r$ is maximised
in the `Zeno regime' $\Omega_{\rm eff}^2 \ll R^2$. It is
always possible to enter this regime without affecting the
light shift by reducing $\Omega_1$ at fixed values of $\Omega_2$
and $\Delta_1$. From (\ref{bright2}) and (\ref{dark}) we then
obtain
\beq
r(\Delta'=Z) = \frac{Z^2 + \gamma^2}
{ ( \alpha R/2 + \gamma ) \gamma }.
\eeq
To maximise $r$, one should reduce $R$ as much as possible,
subject to the constraint $\Delta'=Z$. This means that, for given $Z$,
the value of $R$ is limited by the available laser power:
$R = 4 Z^2 \Gamma/\Omega_2^2$, so
\beq
r(\Delta'=Z) = \frac{Z^2  + \gamma^2}
{(2 \alpha \Gamma Z^2 /\Omega_2^2 + \gamma) \gamma}.
\label{result}
\eeq
This is the same result as (\ref{rinf}). Therefore if
$\Omega_1$ is reduced sufficiently to enter the Zeno regime,
then for laser linewidths satisfying $\gamma \ll \Omega_2^2 / \Gamma$,
the value of $r$ is the same at
$\Delta_1 = \Omega_2^2 / 4Z$ (where the D system is tuned to dark
resonance) as when
$\Delta_1 \rightarrow \infty$.

\subsection{Discussion}

\begin{figure}[t]
\centerline{\resizebox{!}{6 cm}{\includegraphics{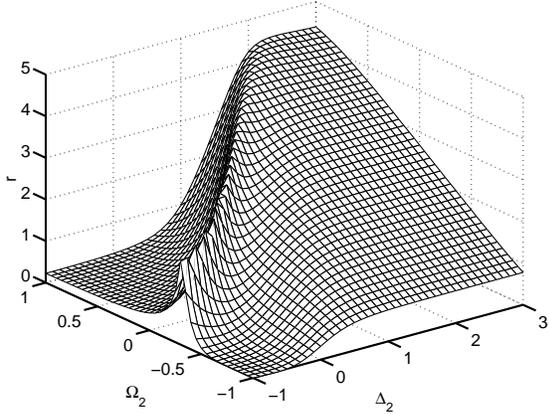}}}
\caption{Discrimination ratio $r$ for the case of
two $\Lambda-$systems with the same coupling constants,
and 2-photon resonance conditions of frequency separation $Z$. The surface
shows $r$ as a function of 
$\Delta_2$ and $\Omega_2$
for the case $Z=0.2$, $\gamma=0.001$,
%$\Omega_2 = 0.5,1,2,4,8$, 
and small $\Omega_1$, in units where $\Gamma=1$.
%The crosses
%mark the points where $\Delta' = Z$. 
All scales are logarithmic, marked in powers of 10.
Note that the range of validity of the
approximate equation (\ref{full}) is such that it gives the same results (i.e.
no discernible difference in this surface plot) as the exact
equation (\ref{rho33exact}), even where $\Delta_1$ is small.} \label{f:6}
\end{figure}

The ratio $r = \rho_{33}(\delta) / \rho_{33}(\delta-Z)$ is plotted
in figure \ref{f:6}
as a function of pump laser parameters $\Omega_2$, $\Delta_2$, for
the example case of $Z=0.2 \Gamma$, $\gamma=0.001 \Gamma$ and small
$\Omega_1$.   
The ridge observed in the surface
corresponds to the condition $\Delta' = Z$, with a slight offset
owing to the displacement of the absorption minimum remarked in the
caption to figure \ref{f:2} (see below). Each line of 
$r$ at constant $\Omega_2$
has a local maximum at the ridge, and then tends to this same maximum $r$
at large $\Delta_2$. This is the basic behaviour predicted by equations
(\ref{rinf}) and (\ref{result}). 
A wider numerical exploration indicated that the value given by
(\ref{rinf}) and (\ref{result}) is always close to the maximum $r$
when $\gamma$ is small enough to allow
good discrimination ($r \gg 1$). 

Equations (\ref{rinf}) and (\ref{result})
are among the central results of this
paper. We had expected that
arranging the special case where
the light-shift $\Delta'$ matches the offset $Z$
would provide an especially good discrimination, as
quantified by the ratio $r$. However,
although we find that this case does
provide the maximum $r$ at given $Z$, $\gamma$ and $\Omega_2$, we
find that the same value of $r$ is also available
when $\Delta' \ne Z$ by using a large detuning.
Therefore the EIT can be useful to increase the
rate of signal acquisition, but it does not provide an improved
discrimination of the two resonances in the atom.
Hence the title of
this paper is a misnomer for the case considered here:
the most important feature
is the presence of the bright resonance, not the dark
resonance. This could be called quantum state discrimination
by `EIO', that is, electromagnetically-induced {\em opacity}.

At small $\Omega_2$ and $\gamma$,
$r$ increases as $\Omega_2^2$ and does not depend on $Z$,
while at large $\Omega_2$ it saturates to $r \rightarrow Z^2 / \gamma^2 +1$.
The latter result is exactly the same as
equation (\ref{r1}) for single-photon excitation limited
by laser linewidth, if for given $Z$ we compare the summed laser linewidths in
the 2-photon case with the single laser linewidth in the single-photon case.
This is owing to the Fano lineshape becoming Lorentzian
when the laser linewidth dominates its FWHM. The surprising
feature is that choosing laser parameters in order to get
a non-Lorentzian Fano profile, with its apparently useful
sub-Lorentzian behaviour near $\delta=0$, in fact can only make
matters worse at given laser linewidth and intensity.

Close inspection of the numerical results reveals a further detail.
This is that for a strong pump beam, the optimal
detuning is larger than that which leads to $\Delta'=Z$, and
a slightly increased $r$ is available. This is owing to the fact
that for finite $\gamma$ the minimum absorption is displaced
from $\delta=0$, as shown in figure \ref{f:2}. We find that this
offset is given by $2\gamma \Delta_1 /(\alpha \Gamma + 4 \gamma)$, in
agreement with \cite{kofman1997}. An increase in $\Delta_1$ reduces
the light shift and hence allows the D manifold to be closer to the
minimum when the B manifold is at the peak. 

To summarize, in the case of two $\Lambda$-systems of the
same coupling constant but different energy level separations,
we find that the highest value of $r$ is obtained
both at $\Delta' = Z$, and at large $\Delta_2$. Going to large $\Delta_2$
has the disadvantage that the rates get small, so the system
is more sensitive to drifts and other line-broadening
mechanisms. Therefore the optimum conditions are, for given
$Z$, $\gamma$:
\begin{eqnarray}
\Omega_2 && \mbox{as large as possible} \\
\Delta_2 &=& \frac{ \Omega_2^2 }{4 Z}\left( 1 + \frac{\gamma \Omega_2^2}{2 \Gamma Z^2} \right) - Z 
\label{D2}\\
\Omega_1 & \ll & \max \left( \frac{Z \Gamma}{\Omega_2}, \;
\frac{\Delta \gamma}{\Omega_2} \right)  \label{small1}
\end{eqnarray}
where in (\ref{D2}) we have included an adjustment for the displaced minimum, and
the condition (\ref{small1}) is to avoid power-broadening of the bright resonance.
Equation (\ref{result}) shows also that smaller laser linewidth is always
advantageous to increase $r$, whereas $r$ saturates as a function
of $Z$, ceasing to increase significantly with $Z$ once $Z$ is large
compared to $\Omega_2 (\gamma / 2 \Gamma)^{1/2} $.

These conclusions are valid when the laser linewidth is caused
by, or is equivalent to, phase diffusion. If other sources
of noise, such as jitter and drift, dominate
(with a non-Lorentzian frequency distribution)
then the evaluation of $r$ has to be reconsidered. In some
circumstances it is appropriate to take average values
of $\rho_{33}^{\rm dark}$ and $\rho_{33}^{\rm bright}$,
using equations (\ref{bright}) and (\ref{dark}) averaged
over the relevant laser frequency distribution. In certain
cases the dark resonance can allow a much greater discrimination
than would be obtained using narrow single-photon transitions
driven by lasers with the same frequency distribution.

\section{Laser cooling of a trapped atom} \label{s:cool}

Laser cooling of a trapped 3-level atom using narrow two-photon resonances has
been discussed by various authors, see
ref. \cite{lindberg1986,marzoli1994} and references therein for 
a general discussion. We will examine the specific case of
using the bright resonance (and accompanying dark resonance) for continuous cooling;
this was considered by
\cite{lindberg1986,marzoli1994,reiss1996,reiss2002,morigi2000,morigi2003}.

Using the formulation as given by \cite{lindberg1986}, we obtain the
steady state solution 
for the motional density matrix $\rho_m$ of a trapped atom or ion in
the Lamb-Dicke limit.  
Expanding the master equation to lowest order in the Lamb-Dicke parameters
$\eta_1, \eta_2$ (associated with the laser excitation
on transitions $1 \leftrightarrow 3$, $2 \leftrightarrow 3$ respectively), the
solution is found to be a thermal state $\rho_m = \sum (1-q) q^n \ket{n}\bra{n}$ where
$q = A_+ / A_-$ is the ratio of heating- to cooling-rate coefficients.
The rate coefficients $A_{\pm}$ are given by 
\begin{eqnarray}
A_{\pm} & = & \left(\Gamma_1 \alpha_1 \eta_1^2 + \Gamma_2 \alpha_2
  \eta_2^2\right) \rho_{33} \nonumber \\
& & + \mbox{Re} \left\{ \mbox{ Tr} \left[ 2 V (L_0 \pm i \nu)^{-1} V
  \rho \right]\right\} 
\end{eqnarray}
where $\alpha_1, \alpha_2$ are
coefficients describing the angular distribution of spontaneously emitted photons (e.g. $\alpha=1/3$
for isotropic emission), $\rho_{33}$ is the internal upper state population in steady state with motional
effects ignored, i.e. as given by (\ref{rho33exact}), $V$ is the internal-state part of the laser--atom
interaction which corresponds to first sideband excitation:
\[
V = \eta_1 \frac{\Omega_1}{2}(\ket{3}\bra{1} - \ket{1}\bra{3})
+ \eta_2 \frac{\Omega_2}{2}(\ket{3}\bra{2} - \ket{2}\bra{3}),
\]
and $L_0$ is the zeroth order Liouville operator acting on the internal state, defined
such that the master equation
$\dot{\rho}= L_0 (\rho)$ gives precisely the OBEs for the semi-classical treatment of
a free atom, as given in (\ref{1})--(\ref{6}).

This situation may be compared with the selective excitation which is the main subject of this
paper. 
Let $\omega_z$ be the vibrational frequency of the given atom in the (assumed harmonic) trap.
Efficient cooling, and low steady-state temperature, is obtained when the cooling
rate $A_-$ is high and the heating rate $A_+$ is low. This requires strong excitation
of the first red sideband at $\omega_0 - \omega_z$
while avoiding excitation of the carrier and the first blue
sideband, at $\omega_0$ and $\omega_0 + \omega_z$ respectively, where $\omega_0$ is the
centre of some resonance feature in the excitation spectrum of a free atom---in
our case, the bright resonance. 
The energy level structure is akin to that of fig. 1b rather than 1a, since the
ladder of vibrational energy levels leads to an infinite set of
$\Lambda$-systems. 
To obtain an enhancement from EIT, the lasers should be blue detuned,
i.e. $\Delta_{1}, \Delta_{2} > 0$.
The frequency difference $Z$ considered in section \ref{s:bright}
corresponds to the vibrational energy $\omega_z$. The selectivity parameter
$r$ discussed in section \ref{s:bright} corresponds to $1/q$.
Just as we suspected that we
might observe large selectivity $r$ when $Z=\Delta'$, we now investigate whether
we observe an especially low $q$ when $\omega_z = \Delta'$.

\begin{figure}[h]
\centerline{\resizebox{!}{6 cm}{\includegraphics{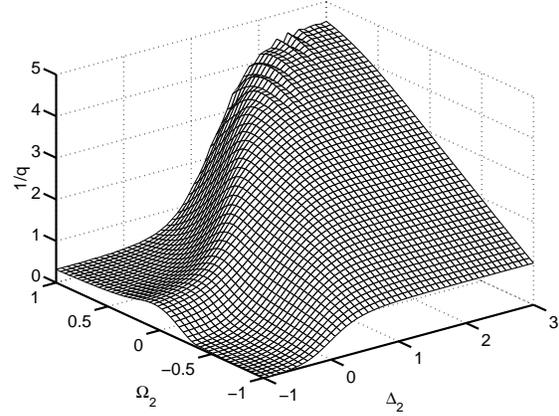}}}
\caption{Cooling/heating ratio $1/q$ for the case of laser cooling of a trapped three-level
atom using the bright resonance. The surface
shows $1/q$ as a function of 
$\Delta_2$ and $\Omega_2$
for the case $\omega_z=0.2$, $\gamma=0.001$,
and small $\Omega_1$, $\eta_1$, $\eta_2$, in units where $\Gamma=1$.
All scales are logarithmic, marked in powers of 10. (The small irregular ripples
at large $1/q$ are a numerical artifact.)} \label{f:7}
\end{figure}

Figure \ref{f:7} shows $1/q$ for the case of laser cooling, for the same parameters
as were chosen in figure \ref{f:6} for the case of $r$ and selective excitation. 
The two sets of results are broadly similar. The main difference is that the
ridge (i.e. high value
of $1/q$, giving low temperature) produced by the `EIT condition' $\omega_z = \Delta'$ is now
lower and broader, compared to the ridge in $r$ in fig. \ref{f:6}.
This is because we now have many $\Lambda$ systems, and the
heating coefficient $A_+$ is produced both by the carrier and the blue sideband
excitation: the dark resonance can suppress one or other of these, but not both.
As a result, the lines of $1/q$ at constant $\Omega_2$ show no
local maximum as a function of $\Delta_2$. 
$q$ (and hence the steady-state temperature)
falls monotonically as a function of pump laser detuning. 

Although the EIT condition does not produce the lowest steady-state temperature $T_0$,
for given values of pump laser intensity and trap frequency, it can be useful
for other reasons. For example it was shown in \cite{morigi2000} that it produces 
a high ratio $A_- / T_0$ of cooling rate to steady-state temperature, and
permits cooling of motion in all directions to the same $T_0$.

\acknowledgements
We thank Dr Giovanna Morigi and Dr J\"{u}rgen Eschner for helpful
discussions on laser cooling using EIT, and
Dr David McGloin and Dr David Lucas 
for comments on the manuscript. This work was supported by the EPSRC,
ARDA (P-43513-PH-QCO-02107-1) the Research Training and Development and Human Potential
Programs of the European Union, and the Commonwealth Scholarship and Fellowship Plan.

\section{Appendix}

Here we present the solution of the OBEs for a 3-level $\Lambda$-type system.

The solution for $\rho_{13}$ can be extracted by a standard matrix inversion method,
see for example \cite{book:scully}, where the case $\Delta_2=0$ is treated in full.
We are interested here in $\rho_{33}$ so we present this quantity.

The solution for $\gamma=0$ has been presented by various authors,
see for example \cite{janik1985} whose notation is close to ours.
The solution for general $\gamma$ was discussed in \cite{brewer1975,lindberg1986}
and is closely related to the
ladder system discussed in \cite{whitley1976}. However the expressions in
these works are even
more lengthy and obscure than those given below; we require the simplest
form possible. 

In order to simplify the expressions without much loss
of generality, we assume
$\Gamma_{13} = \Gamma_{23}$ (this is valid when the lasers'
linewidths are equal, and approximately valid when they are unequal but
small compared to $\Gamma$).

In this case, the steady state value
of $\rho_{33}$ is as given in (\ref{rho33exact}), with the
coefficients in the denominator as follows:
\begin{eqnarray}
c_0 &=& (\Omega_1^2 + \Omega_2^2)^2 Y + 16 \delta^2 \Gamma_{13}^2 Y  \nonumber\\
  & &  + 4 \delta^2 \Omega_1^2 \Omega_2^2 (6 \Gamma_{13} - (\Gamma_1 + \Gamma_2))
 \nonumber \\
&& 
+ 16 \delta^2 (\Gamma_2 \Omega_1^2 \Delta_2^2 + \Gamma_1 \Omega_2^2
  \Delta_1^2)
\nonumber \\
& & -8 \delta (\Delta_1 \Gamma_1 \Omega_2^4 - \Delta_2 \Gamma_2 \Omega_1^4)   \label{c0exact}
\end{eqnarray}
where $Y \equiv \Gamma_2 \Omega_1^2 + \Gamma_1 \Omega_2^2$,
\begin{eqnarray}
\lefteqn{c_1 = 
2 \left(\Omega_1^2 + \Omega_2^2\right) \left(4\Gamma_{13} Y + 3 \Omega_1^2 \Omega_2^2\right) } \\
&&+ 4 \frac{\Omega_1^2 \Omega_2^2}{\Gamma_{13}}\left(\Gamma_1 \Delta_1^2 + \Gamma_2 \Delta_2^2 
+ (\Gamma_1+\Gamma_2) \Delta_1 \Delta_2\right), \nonumber
\end{eqnarray}
and
\begin{eqnarray}
c_2 &=& 8 \left[ 2 \Gamma_{13}^2 Y + 3 \Gamma_{13} \Omega_1^2
  \Omega_2^2 \nonumber\right. \\
 & &\left. + 2\left(\Delta_2^2 \Gamma_2 \Omega_1^2 + \Delta_1^2 \Gamma_1
  \Omega_2^2\right) \right]. 
\end{eqnarray}

Equation (\ref{c0exact}) can also be written:
\begin{eqnarray}
c_0 &=& 16 \Omega_2^2 \Gamma_1 \Delta_1^2 \left( \delta -
  \frac{\Omega_2^2}{4\Delta_1}\right)^2 \nonumber \\
  & & + 16 \Omega_1^2 \Gamma_2 \Delta_2^2 \left( \delta -
  \frac{\Omega_1^2}{4\Delta_2}\right)^2  
\nonumber \\
&&
+ 16 \delta^2 \Gamma_{13}^2 Y + 
4 \delta^2 \Omega_1^2 \Omega_2^2 (6 \Gamma_{13} - (\Gamma_1 + \Gamma_2))   \nonumber \\
&&
+ \Omega_1^2 \Omega_2^2 \left(Y + \left(\Omega_1^2 + \Omega_2^2\right)(\Gamma_1 + \Gamma_2)\right).
\end{eqnarray}
This form is useful in order to clarify where the resonances are, and to derive
equation (\ref{c0}).

\subsection{Degenerate $\Lambda$-systems}

Here we briefly discuss the case of two degenerate $\Lambda$-systems,
but where discrimination is still possible
because of a difference in coupling constants.

We adapt the notation so that now the parameters $\Omega_i$
refer to manifold D, and we define $C_i = \Omega_i^{\rm B} /\Omega_i$,
$i=1,2$
where $\Omega_i^{\rm B}$ are the Rabi frequencies in manifold B.
The maximum $r$ occurs either when system B is tuned to bright
resonance, or when system D is tuned to dark resonance. The latter
case is only relevant when $\gamma$ is very small, and then
$r$ is a ratio of two very small excitation rates. We will
concentrate on the case where $\gamma$ is somewhat larger, and
then it is best to tune B to bright resonance. We then have
$r = \rho_{33}^{\rm B} / \rho_{33}^{\rm D}$ where $\rho_{33}^{\rm B}$ is given
by equation (\ref{bright2}):
\beq
\rho_{33}^{\rm B} =
\frac{C_1^2 C_2^2 \Omega_{\rm eff}^2 / 2 \Gamma_1 C_1^2}{\alpha C_2^2 R/2 +
\gamma}.
\eeq
The symbols $\Omega_{\rm eff}$, $R$ refer to their values in
system D, and we assume the decay rate $\Gamma_1$ is enhanced in
system B, compared to D, by $C_1^2$. We have also assumed the Zeno
regime in order to avoid power-broadening.

The situation in manifold D is given by equation (\ref{fullapp})
at $\delta = \Delta'_{\rm B} - \Delta'_{\rm D} = (C_2^2 - 1) \Delta'$, hence:
\beq
\rho_{33}^{\rm D} =
\frac{ \Omega_{\rm eff}^2 ( \alpha R C_2^4 / 2 + \gamma )/ 2 \Gamma_1}
{ (C_2^2 - 1) \Delta'^2 + (\alpha R/2 + \gamma)^2 },
\eeq
where we assumed (\ref{smallgam}). The largest values of $r$ are
obtained at high detuning, such that $R \ll \gamma$, where we find
\beq
r(\Delta_1 \rightarrow \infty) = \left( C_2(C_2 -1)\Delta'
\right)^2 / \gamma^2 + 1.
\eeq

\bibliographystyle{unsrt}

\end{document}